\documentclass[preprint,3p,12pt]{elsarticle}

\usepackage{amssymb}
\usepackage{amsmath}
\usepackage{graphicx}
\usepackage{xcolor}
\usepackage[colorlinks=true,pdfa=true]{hyperref}

\journal{PoS: 38th ICRC 2023}

\begin{document}

\begin{frontmatter}

{

\title{CR-ENTREES - Cosmic-Ray ENergy TRansport in timE-Evolving astrophysical Settings}

\author[a]{Anita Reimer\corref{cor1}}
\cortext[cor1]{Corresponding author}
\ead{anita.reimer(at)uibk.ac.at}
\author[a,b,c]{Lukas Merten}
\author[a]{Margot Boughelilba}
\author[a]{Paolo Da Vela}
\author[d]{Serguei Vorobiov}
\author[d]{Jon Paul Lundquist}

\affiliation[a]{organization={Universitaet Innsbruck, Institut fuer Astro- und Teilchenphysik},
            city={Innsbruck},
            postcode={6020}, 
            country={Austria}}
\affiliation[b]{organization={Theoretical Physics IV, Plasma Astroparticle Physics, Faculty for Physics and Astronomy, Ruhr University Bochum},
            city={Bochum},
            postcode={44780}, 
            country={Germany}}
\affiliation[c]{organization={Ruhr Astroparticle and Plasma Physics Center (RAPP Center)},
            city={Bochum},
            postcode={44780}, 
            country={Germany}}
\affiliation[d]{organization={Center for Astrophysics and Cosmology (CAC), University of Nova Gorica},
            city={Nova Gorica},
            postcode={SI-5000}, 
            country={Slovenia}}

\begin{abstract}

In order to understand observable signatures from putative cosmic-ray (CR) sources in-source acceleration of particles, their energy and time-dependent transport including interactions in an evolving environment and their escape from source have to be considered, in addition to source-to-Earth propagation.

We present the code CR-ENTREES (Cosmic-Ray ENergy TRansport in timE-Evolving astrophysical Settings) that evolves the coupled time- and energy-dependent kinetic equations for cosmic-ray nucleons, pions, muons, electrons, positrons, photons and neutrinos in a one-zone setup of (possibly) non-constant size, with user-defined particle and photon injection laws. All relevant interactions, particle/photon escape and adiabatic losses are considered in a radiation-dominated, magnetized astrophysical environment that is itself evolving in time. Particle and photon interactions are pre-calculated using event generators assuring an accurate interactions and secondary particle production description. We use the matrix multiplication method for fast radiation and particle energy transport which allows also an efficient treatment of transport non-linearities due to the produced particles/photons being fed back into the simulation chain. 

Examples for the temporal evolution of the non-thermal emission from AGN jet-like systems with focus on proton-initiated pair cascades inside an expanding versus straight jet emission region, are further presented.

\end{abstract}

}

\end{frontmatter}

\section{Introduction}

With more than 60\% of all the detected sources in the $\gamma$-ray sky belonging to the class of jetted active galactic nuclei (AGN), and their proposed contribution to the PeV neutrino and ultra-high energy cosmic-ray (UHECR) sky, a deep exploration of the multi-messenger nature of this source class is central. Where and how are the high-energy messengers produced? How are the charged particles accelerated to such extreme energies and how can they escape the magnetized jet environment? What is the overall composition of the jet? A tool that supports the investigation of these questions is presented here: CR-ENTREES, a code for fully time-dependent \underline{C}osmic-\underline{R}ay \underline{EN}ergy \underline{TR}ansport in tim\underline{E}-\underline{E}volving astrophysical \underline{S}ettings.

CR-ENTREES is used as the base code for the heavy nuclei propagation code of \cite{Merten} (see this proceedings).

\section{Propagation Physics}

CR-ENTREES solves the following nonlinear system of coupled (integro-differential) Fokker-Planck transport equations:
$$
\partial_t F_N  + \dot F^{\rm esc}_N + \partial_E[(\dot E_{\rm loss}F_N)]+\dot F^{\rm dec}_N = Q^{\rm inj,pr}_N
$$
$$
\partial_tF_{\mu,\pi,K}  + \dot F^{\rm esc}_{\mu,\pi,K} + \partial_E[(\dot E_{\rm loss}F_{\mu,\pi,K})]+\dot F^{\rm dec}_{\mu,\pi,K} = \dot F^{p\gamma;h}_{\mu,\pi,K}
$$
$$
\partial_tF_e  + \dot F^{\rm esc}_e + \partial_E[(\dot E_{\rm loss}F_e)]= Q^{\rm inj,pr}_e+\dot F_e^{\gamma\gamma}+\dot F_e^{p\gamma}
$$
$$
\partial_tF_\gamma  + \dot F^{\rm esc}_\gamma + \dot F_\gamma^{\gamma\gamma}= \dot F_\gamma^{\rm em}+\dot F_\gamma^{p\gamma;h}
$$
Here,
$F_{\rm X} = F_{\rm X}(E,t)$ is the energy (E) and time (t) dependent density of the cosmic-ray (CR) nucleons $X=N$, electrons and positrons $X=e$, muons, pions, kaons $X=\mu, \pi, K$, respectively, and photons $X=\gamma$. $Q^{\rm inj,pr}_{N,e} = Q^{\rm inj,pr}_{N,e}(E,t)$ describes the source function of the primary nucleons (N) and pairs (e), and
$\dot E_{\rm loss}=\dot E_{\rm loss}(F_\gamma(\epsilon,t),B(t); E,t)$ the continuous loss processes that potentially depend on the (possibly evolving) magnetic field strength $B(t)$ and a energy ($\epsilon$) and time dependent target radiation field. Injection of secondary particles and photons produced in hadronic (h) and electromagnetic (em) particle-photon, photon-photon and particle-field interactions are considered with the density rates
$\dot F_e^{p\gamma}=\dot F_e^{p\gamma}(F_\gamma(\epsilon,t); E,t)$, $\dot F^{p\gamma;h}_{\mu,\pi,K}$,
$\dot F_e^{\gamma\gamma}=\dot F_e^{\gamma\gamma}(F_\gamma(\epsilon,t); E,t)$,
$\dot F_\gamma^{\gamma\gamma}=\dot F_\gamma^{\gamma\gamma}(F_\gamma(\epsilon,t); \epsilon,t)$,
$\dot F_\gamma^{p\gamma}=\dot F_\gamma^{p\gamma}(F_\gamma(\epsilon,t); \epsilon,t)$.
The rigorous treatment of a time-evolving emission volume, magnetic field and target radiation field within the present setup causes the non-linearity of this transport equation system.
The user of CR-ENTREES can choose to consider the continuous loss processes inverse Compton scattering, Bethe-Heitler pair production and synchrotron radiation of all charged particles, and the catastrophic loss processes photomeson production, particle decay and escape. For a conical jet setup adiabatic losses are taken into account as well. We note that in the absence of nucleons the system becomes identical to a Synchrotron-Self Compton model.

\section{The Model}

While CR-ENTREES (thanks to its modular implementation) is flexible to be adapted to a large range of geometrical setups where CR transport occurs, its first purpose is to describe a moving homogeneous one-zone emission region within a straight or conical outflow/jet. Here, CR transport is treated in the co-moving frame. Hence, all input parameters (except the outflow speed - see below) are considered in this frame.

\subsection{Input: Characterizing geometry and environment}

The moving (with given constant speed $\beta_{\rm J} c$) spherical emission region of fixed or time-evolving (in case of a conical outflow) radius $R$ contains a homogeneous magnetic field of strength $B$ and (assumed) isotropic target radiation field for particle-photon interactions. Magnetic field evolution is currently implemented with a $B(t)\propto t^{-1}$ scaling (with $t$ the co-moving propagation time) in case of a conical jet setup. The target photon field density distribution is discretized on a fixed 161 log-equal spacing energy grid in the range $10^{-10 ... 6}$eV. A (diluted) blackbody with given temperature $T$ or power-law spectra (normalized using a given energy density) with up to 2 break energies (and corresponding power-law indices) can be chosen to fill this target field. Alternatively, this field can be filled for each energy bin by the user. The total target radiation field is then determined from the internal (jet) radiation field (calculated in each time step) and the user-defined target field. 

\subsection{Input: Characterizing particle injection}

Power-law spectra (of possibly exponential cutoff) of any particle type (or photon) within a user-defined energy range and index are injected into the simulation chain. These injection spectra (and all particle spectra during propagation)
are discretized on a fixed 300 log-equal spacing energy grid in the range $10^{-3 ... 12}$GeV. Currently, up to 2 particle populations can be injected. For their normalization the number ratio of these populations and the particle-to-field energy density ratio must be provided by the user.

All input parameter values are provided by the user in a dedicated steering file. 

\subsection{Energy loss processes and secondary particle production}

We use Monte Carlo event generators (photomeson production and decay processes: \cite{SOPHIA}; Bethe-Heitler pair production: modified version of \cite{Ray96}; inverse Compton scattering and photon-photon pair production: \cite{Ray86}) to pre-calculate the yields and interaction rates (which are stored in HDF5-files) of each interaction type, and discretized on a fixed 300 log-equal spacing energy grid in the range $10^{-3 ... 12}$GeV for a range of energies for the target particle/photon. The corresponding yield and interaction rate for the (assumed isotropic) target radiation field in each time step are then determined by convolving over this target field.
Synchrotron radiation yields and corresponding loss rates (using pitch-angle averaged terms) of all produced charged particles (i.e., charged pions, muons, kaons and electrons/positrons) are calculated following \cite{Pacho}, \cite{Brainerd}, and \cite{Rybicki} (for the self-absorption process).
Note that particles suffering from continuous losses in combination with catastrophic losses on a fixed energy grid require a dedicated numerical treatment to let these particles move down this energy grid from high to low energies.

The adiabatic loss time scale for particles of Lorentz factor $\gamma$ in a conical jet of opening angle $0.26/\Gamma_{\rm J}$ \cite{Pushkarev}, with $\Gamma_{\rm J} = (1-\beta_{\rm J}^2)^{-0.5}$ the bulk Lorentz factor of the moving emission region, is calculated to $t_{\rm ad}=\frac{\gamma^2}{\gamma^2-1} t \left (1-\frac{R_0}{R(t)}\right )^{-1}$ with $R_0$ the size of the emission region at propagation time $t=0$. Note the decrease of the adiabatic loss rate with increasing propagation time.

Finally, neutral particles (including photons) escape on a time scale $t_{\rm esc,n}=\frac{3}{4} R(t)/c$, while for the charged particles' escape time scale $t_{\rm esc,c}=\eta t_{\rm esc,n}$ with $\eta\geq 1$, is used.

All particle energy, adiabatic and escape losses are tracked as well throughout the entire simulation chain to allow for the verification of energy conservation in each simulation time step.

\subsection{Propagation method}

We propagate $\gamma$-ray, proton, neutron, electronic pair, muon, pion, kaon, muon and electron neutrino populations, discretized on a fixed 300 log-equal spacing energy grid in the range $10^{-3 ... 12}$GeV, and lower energy photons on a fixed 300 log-equal spacing energy grid in the range $10^{-18 ... -3}$GeV, using the matrix multiplication method of \cite{Ray86}, \cite{Ray93}, \cite{Ray96} in the framework of CR transport. Here, transfer matrices are created 
from the aforementioned yields and interaction probabilities which describe the change after a given time step $\delta t$ of the density of a given particle type upon all the interactions pre-set by the user. Such explicit integration scheme, while extremely fast, requires to use time steps that are smaller than the smallest time scale of the system (the Courant-Friedrichs-Levy condition). Still, with the matrix doubling method of \cite{Ray93} applied, we found correct and stable results also for somewhat larger time steps. Here, energy conservation is verified after each time step.

\subsection{Code output}

CR-ENTREES' output encompasses the co-moving density of all propagated particle types after the pre-set propagation time, and interaction rates of all chosen processes, on the aforementioned energy grid. Post-processing of the output density within the emission volume to transform into the observer frame is carried out outside the CR-ENTREES framework.


\section{Examples}

As an example for the use of CR-ENTREES we show here the development of a pair cascade initiated by a cosmic-ray proton population inside a magnetized, relativistically moving emission region along a straight, and a conical jet. In our example model the initial size of the comoving emission region amounts to $R_{\rm emi}=3\times 10^{16}$cm, located at a distance of $10^{17}$cm away from the central engine, and moves with a speed that corresponds to a bulk Lorentz factor of $\Gamma_{\rm J}=22.4=D$ (where $D$ is the Doppler factor). We instantaneously inject a relativistic electron population, that follows a $\propto\gamma_e^{-2}$ particle energy distribution between Lorentz factors $\gamma_e=1$ and $\gamma_e=100$, into the magnetized (with initial magnetic field strength of $10$G) emission region. These electrons build up a synchrotrotron radiation field, which has its power peak in the X-ray domain (see Figs. 1, 2), resembling an HBL-like synchrotron spectrum to some extend. Cosmic-ray protons, injected along with the electron population with the same spectral index but in an extended particle energy range from $\gamma_p=1$ to $\gamma_p=10^9$, initiate then, via proton-photon interactions in this evolving synchrotron radiation field, a pair cascade, which we follow with time steps of $\delta t = 10^4$s.
With an initial particle-to-field energy density ratio of $10$, and a proton-to-electron injection number density ratio of $10^5$, the normalization of the two injected particle populations are set. In order to focus on the cascade development, we switch off proton synchrotron radiation in these simulations. All other processes, including charged particle escape (on a time scale $t_{\rm esc,c}=3/4 R_{\rm emi}/c$) are kept active. 

Figure~1 shows the evolution of the broadband photon SED from a straight jet moving emission region in the galaxy frame at times $t_{\rm obs}\cdot D=10^4$s, $2\times 10^4$s, $3\times 10^4$s, $10^5$s, $2\times 10^5$s, $8\times 10^5$s (black lines) and $t_{\rm obs}\cdot D=1.8\times 10^6$s, $2.8\times 10^6$s, $5.9\times 10^6$s, $8.9\times 10^6$s (grey lines) 
after the injection. The synchrotron radiation field, target for particle-photon interactions, reaches its maximum power within less than one dynamical time scale, its decline is much slower. Figure~3 shows the correspondingly strongly asymmetric X-ray light curve (dashed red line). Also the subsequently developing pair cascade responds with a fast flux increase, and a slower decrease past its power maximum (see the $\gamma$-ray light curves in Fig.~3). The 1~PeV neutrino light curve (violet line in Fig.~3) responds even slower: Its power maximum is reached within less than a day (galaxy frame), its decline starts clearly after the decrease of the photon cascade's emission.

We then study such proton-initiated pair cascade development, using the same model parameters as above, in a moving emission region within now a conical jet of opening angle $\sim 0.66^{\rm o}$. The adiabatic expansion of the emission region leads in this case  to further energy losses. At the same time, the corresponding increase of the comoving escape time scales (from $7.5\times 10^{5}$s at injection time to  $3.7\times 10^{6}$s at $1.5\times 10^7$s after injection at the end of our simulations) as well as the decline of the magnetic field $\propto  t^{-1}$  keeps particles and photons inside the emission region and decreases radiative losses. As a result, the increase and decrease of the target and cascade photon emission is significantly prolonged, as can be appreciated from Figs. 2, 3 of our work: E.g., the duration of the flux enhancement above one order of magnitude below the respective power maximum in the 3 energy bands shown seems a factor of 2-3 longer for the expanding conical jet as compared to the straight jet.

From this study we conclude: The temporal evolution of the size of the emission region, e.g., as a result of the jet's shape, has a significant impact on the resulting multi-messenger light curves.

\begin{figure}
\centering \includegraphics[width=1.\textwidth]{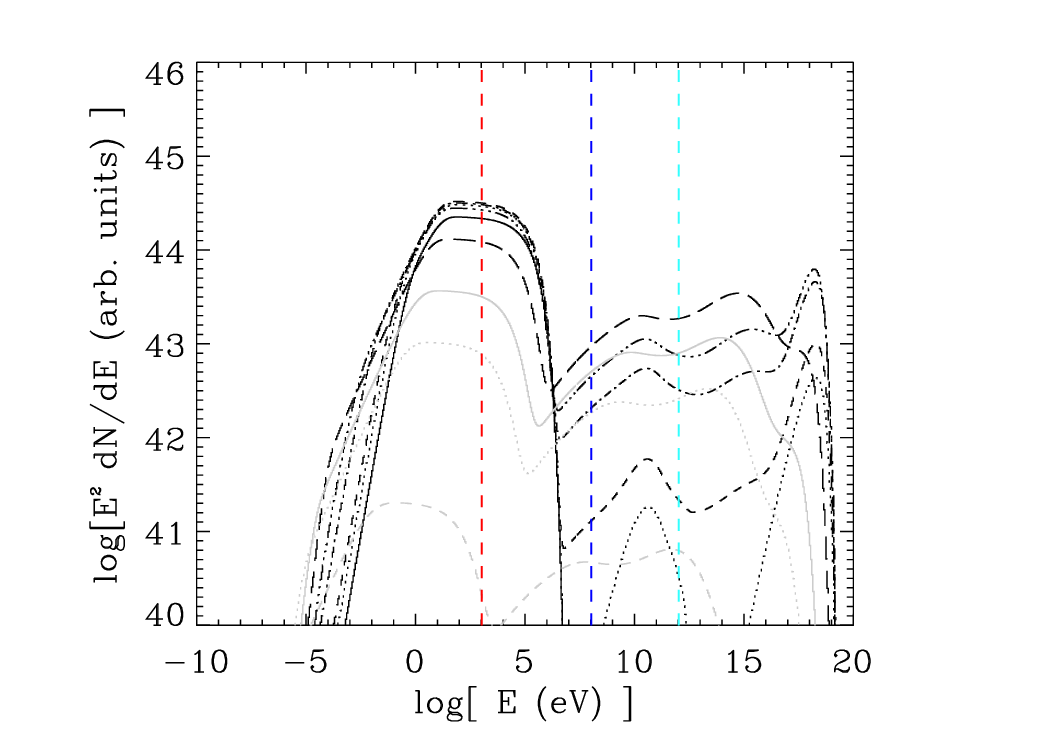}
\caption{Photon SED (host galaxy frame) from the emission region moving along a straight jet with parameters as described in Sect.~4, at (comoving) times $t=10^4$s (solid black line), $2\times 10^4$s  (dotted black line), $3\times 10^4$se (dashed black line), $10^5$s  (dashed-dotted black line), $2\times 10^5$s  (dashed-triple-dotted black line), $8\times 10^5$s (long-dashed black lines) and $t=1.8\times 10^6$s  (solid grey line), $2.8\times 10^6$s  (dotted grey line), $5.9\times 10^6$s  (dashed grey line), $8.9\times 10^6$s (dashed-dotted grey line) after injection. The red dashed line indicates the 1~keV-slice, the blue dashed line the 100~MeV-slice and the pale blue dashed line the 1~TeV-slice through the flux-energy-time data cube. }
\label{fig1}
\end{figure}

\begin{figure}
\centering \includegraphics[width=1.\textwidth]{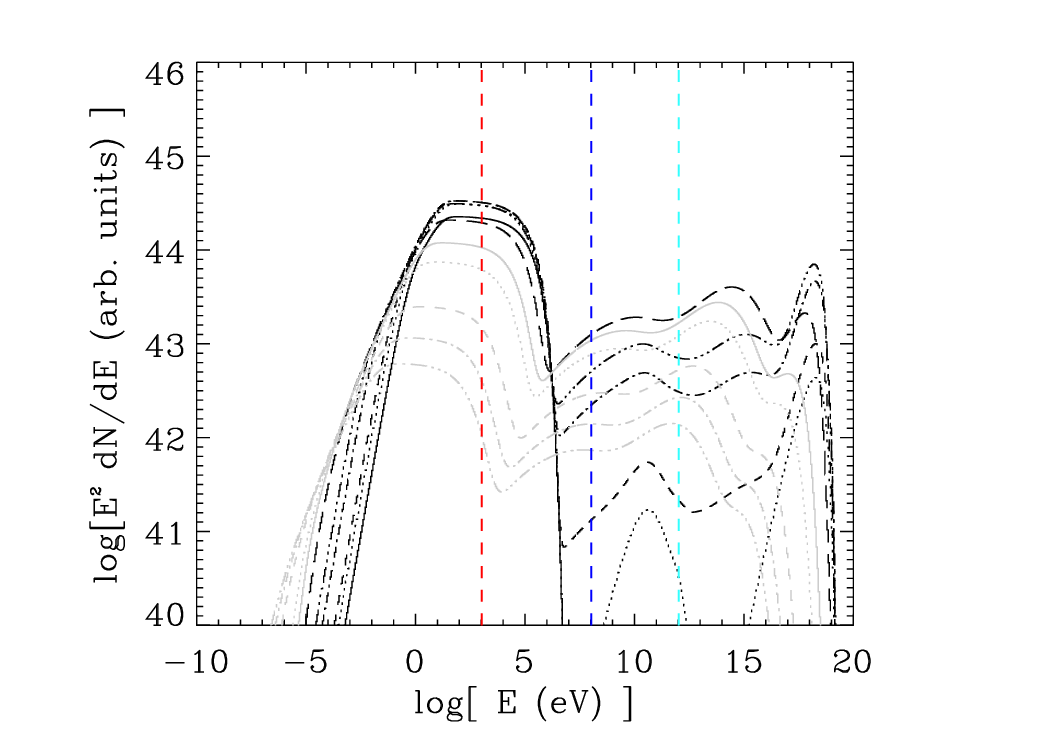}
\caption{Photon SED (host galaxy frame) from the emission region moving along a conical jet with parameters as described in Sect.~4, at (comoving) times $t=10^4$s (solid black line), $2\times 10^4$s  (dotted black line), $3\times 10^4$s  (dashed black line), $10^5$s  (dashed-dotted black line), $2\times 10^5$s  (dashed-triple-dotted black line), $8\times 10^5$s (long-dashed black lines) and $t=1.8\times 10^6$s  (solid grey line), $2.8\times 10^6$s  (dotted grey line), $5.9\times 10^6$s  (dashed grey line), $8.9\times 10^6$s (dashed-dotted grey line),  $1.2\times 10^7$s (dashed-triple-dotted grey line),  $1.5\times 10^7$s (long-dashed grey line)  after injection.  The red dashed line indicates the 1~keV-slice, the blue dashed line the 100~MeV-slice and the pale blue dashed line the 1~TeV-slice through the flux-energy-time data cube.}
\label{fig2}
\end{figure}

\begin{figure}
\centering \includegraphics[width=1.\textwidth]{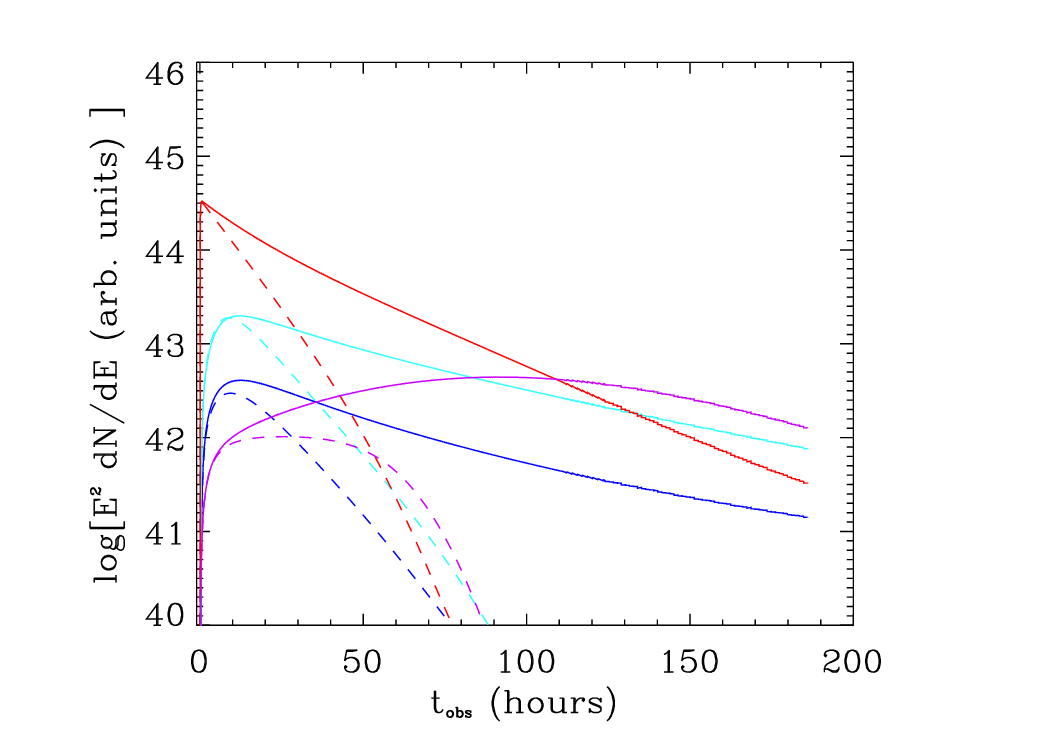}
\caption{Light curves (galaxy frame) from the straight jet (dashed lines) and conical jet (solid lines) emission region taken at photon energies of 1 keV (red lines), 100 MeV (blue lines, 1 TeV (pale blue lines) and at neutrino energy 1 PeV (violet lines).}
\label{fig3}
\end{figure}

\section{Conclusion}
We present CR-ENTREES (Cosmic-Ray ENergy TRansport in timE-Evolving astrophysical Settings), a cosmic-ray energy transport code that solves the nonlinear system of coupled Fokker-Planck transport equations for cosmic-ray nucleons, mesons, leptons, photons and neutrinos. Its flexibility allows to study the resulting multi-messenger emission from particle and photon energy propagation in an environment that is itself changing with time.

In this work CR-ENTREES is used to study the temporal evolution of pair cascades, initiated by the instantaneous injection of relativistic electrons and protons,  in an expanding jet, as compared to a straight jet. We find that the cascade development in an expanding emission region, accompanied with a diluted field environment, leads to a significantly prolonged outburst of multi-messenger emission, as compared to the emission from a fixed-size region in a straight jet.

\section*{Acknowledgements}
Financial support for this project was received from the Austrian Science Fund (FWF) under grant agreement number I 4144-N27 and the Slovenian Research Agency-ARRS (project no. N1-0111). MB has for this project received funding from the European Union’s Horizon 2020 research and innovation program under the Marie Sklodowska-Curie grant agreement No 847476. The views and opinions expressed herein do not necessarily reflect those of the European Commission. LM acknowledges support from the DFG within the collaborative Research Center SFB 1191 "Cosmic Interacting Matters - From Source to Signal".


\begin{thebibliography}{99}
\bibitem[1]{SOPHIA}
A. M\"ucke, R. Engel, J.P. Rachen, R.J. Protheroe, T. Stanev, \emph{Monte Carlo simulations of photohadronic processes in astrophysics},
\href{https://doi.org/10.1016/S0010-4655(99)00446-4}
{\emph{CPC} \textbf{124} (2000) 290}

\bibitem[2]{Ray96}
  R.J. Protheroe, P.A. Johnson \emph{Propagation of ultra high energy protons and gamma rays over cosmological distances and implications for topological defect models}
\href{https://doi.org/10.1016/0927-6505(96)00042-4}
     {\emph{Astroparticle Physics} \textbf{5} (1996) 215}

\bibitem[3]{Ray86}
  R.J. Protheroe, \emph{Effect of electron-photon cascading on the observed energy spectra ofextragalactic sources of ultra-high-energy gamma-rays}
 \href{https://doi.org/10.1093/mnras/221.4.769}
{\emph{MNRAS} \textbf{221} (1986) 769} 

\bibitem[4]{Pacho}
  G.B. Pacholczyk, \emph{Radio Astrophysics}
     Freeman, San Francisco, 1970.

\bibitem[5]{Brainerd}
  J.J. Brainerd, V. Petrosian  \emph{Analytic and Numerical Calculations of Quantum Synchrotron Spectra from Relativistic Electron Distributions}
\href{https://doi.org/10.1086/165588}
     {\emph{ApJ} \textbf{320} (1987) 703}

\bibitem[6]{Rybicki}
  G.B. Rybicki, A.P. Lightman \emph{Radiative Processes in Astrophysics}
  John Wiley \& Sons, 2017.

\bibitem[7]{Pushkarev}
  A.B. Pushkarev, et al, \emph{MOJAVE - XIV. Shapes and opening angles of AGN jets }
\href{https://doi.org/10.1093/mnras/stx854}
     {\emph{MNRAS} \textbf{468} (2017) 4992}

\bibitem[8]{Ray93}
  R.J. Protheroe, T. Stanev \emph{Electron-photon cascading of very high-energy gamma-rays in the infrared background}
\href{https://doi.org/10.1093/mnras/264.1.191}
     {\emph{MNRAS} \textbf{264} (1993) 191}

 \bibitem[9]{Merten}
 L. Merten, P. Da Vela, A. Reimer, M. Boughelilba, J.P. Lundquist, S. Vorobiov, J. Becker Tjus,    \emph{Efficient Modeling of Heavy Cosmic Rays Propagation in Evolving Astrophysical Environments}
{\emph{PoS} (2023)}


\end{thebibliography}
\end{document}